\documentclass[a4paper]{article}

\usepackage{INTERSPEECH2021}
\usepackage{amsfonts, amssymb, amsmath}
\usepackage{mathtools}
\usepackage{array}
\usepackage{bm, bbm}
\usepackage{booktabs}
\usepackage{cite}
\usepackage{enumitem}
\usepackage{graphicx}
\usepackage{hhline}
\usepackage{lipsum}
\usepackage{makecell}
\usepackage{mathtools}
\usepackage{multirow}
\usepackage{siunitx}
\usepackage{tabularx}
\usepackage{tikz}
\usepackage{url}

\DeclarePairedDelimiter\ceil{\lceil}{\rceil}

\newcolumntype{?}{!{\vrule width 0pt}}
\newcolumntype{P}[1]{>{\centering\arraybackslash}p{#1}}
\newcolumntype{M}[1]{>{\centering\arraybackslash}m{#1}}
\DeclarePairedDelimiterX{\distx}[2]{(}{)}{%
  #1\;\delimsize\|\;#2%
}
\newcommand{\dist}{\mathcal{E}\distx}

\newcommand{\setspkagnostic}{\mathbb{G}}
\newcommand{\setspkspecific}{\mathbb{S}^{(k)}}
\newcommand{\setspkspecificpseudo}{\tilde{\mathbb{S}}^{(k)}}

\title{
Personalized Speech Enhancement through\\Self-Supervised Data Augmentation and Purification
}
\name{
Aswin Sivaraman, Sunwoo Kim, Minje Kim}
\address{
Indiana University, Department of Intelligent Systems Engineering, USA}
\email{
\{asivara, kimsunw, minje\}@indiana.edu}

\begin{document}

\maketitle
\begin{abstract}
Training personalized speech enhancement models is innately a no-shot learning problem due to privacy constraints and limited access to noise-free speech from the target user.
If there is an abundance of unlabeled noisy speech from the test-time user, a personalized speech enhancement model can be trained using self-supervised learning.
One straightforward approach to model personalization is to use the target speaker's noisy recordings as pseudo-sources. Then, a pseudo denoising model learns to remove injected training noises and recover the pseudo-sources.
However, this approach is volatile as it depends on the quality of the pseudo-sources, which may be too noisy.
As a remedy, we propose an improvement to the self-supervised approach through data purification. We first train an SNR predictor model to estimate the frame-by-frame SNR of the pseudo-sources. 
Then, the predictor's estimates are converted into weights which adjust the frame-by-frame contribution of the pseudo-sources towards training the personalized model. 
We empirically show that the proposed data purification step improves the usability of the speaker-specific noisy data in the context of personalized speech enhancement. 
Without relying on any clean speech recordings or speaker embeddings, our approach may be seen as privacy-preserving.
\end{abstract}

\section{Introduction}


Speech enhancement is a well-studied research area within signal processing \cite{BollSF79ieeeassp, EphraimY1984spectralamplitude, Gannot98ieeesap} which has experienced significant progress in the past decade due to the pervasiveness of machine learning models and deep neural networks (DNNs) \cite{XuY2014ieeespl, HuangP2015ieeeacmaslp, WeningerF2015lvaica, PascualS2017segan, WangDL2018ieeeacmaslp}. The majority of automated noise suppression algorithms introduced over the years have been geared towards general-purpose (``universal'' or ``speaker-agnostic'') speech enhancement. In this context, denoising models are trained to separate speech from noise without any prior knowledge of the speaker identity or of the noises present. However, given the proliferation of voice-controlled devices (e.g. smart headphones and smart speakers), we anticipate the need for ``personalized speech enhancement'' models which can maximize the enhancement of a single speaker with respect to their unique acoustic environment.

Comprehensive studies of DNN-based speech enhancement or speech separation systems have shown that the generalization power of a model depends on the model's complexity and architecture. For example, a large fully-connected network with $2048$ units and $5$ layers can generalize well to unseen noise sources \cite{ChenJ2016largescale} but may not adapt to unseen test speakers. A long short-term memory cell (LSTM) network achieves the generalization goal in speaker- and noise-agnostic separation tasks \cite{ChenJ2017generalization}. However, it still requires a substantially large network architecture ($1024\times 4$). Other studies have also shown that a mismatch between the training and test input signals may result in highly varied performance unless the model has been exposed to an excessive amount of data \cite{LiuD2014interspeech}. Mismatching factors include the type and loudness of the noise and the characteristics of the speaker. Likewise, in order for a speaker-agnostic generalist model to optimally address the peculiarities of a particular test-time user and their environment, one must both increase the diversity of the training speech and noise corpora and increase the model complexity. 

However, these requirements create a tradeoff, as in personal devices efficient test-time inference is of prime importance due to the often limited resources. In this paper we address this tradeoff by specializing a model. We define a specialist as a smaller model solving a subset of the original problem intended for the generalist. Hence, we can afford a reduction in the overall number of parameters. The benefits of specialization towards speech enhancement have been explored in recent years. The VoiceFilter model informs the speech enhancement model of the estimated speaker-identifying information \cite{WangQ2018voicefilter}. In \cite{KolbakM2017ieeeacmaslp}, speaker- or gender-specific models outperform a generalist model. These studies did not utilize personalization as a manner of reducing model complexity; however, one study extends the idea to a mixture of local expert architecture, where the test-time specificity is identified and then assigned to a few pre-defined specialists, achieving model compression \cite{SivaramanA2020interspeech}.

Another challenge in personalized speech enhancement is that it is not always possible to acquire clean speech data from the test-time user. For example, speech enhancement models within modern-day smart devices might be trained through always-on ambient data collection. This trend is at odds with user concerns regarding privacy and security \cite{rochford2019accessibility}. A recent DNN-based system required as little as five seconds of clean speech data from a test-time speaker in order to convincingly synthesize new utterances out of the previously unseen speaker's voice \cite{JiaY2018neurips}. Breakthroughs such as these may make users reluctant to provide any clean speech recordings to their smart devices. Realistically then, training a personalized model should be viewed as a no-shot machine learning task \cite{ChaoWL2016zeroshotlearning, XianY2018zeroshotlearning}. While eliminating reliance on clean speech recordings from the test-time user may not fully remedy all privacy concerns, we believe speech enhancement models which minimize personal data collection are always desirable from the user's perspective.

In this paper, we take a less intrusive route to achieving personalization by using only noisy data from the test-time speaker. This setup exceeds the scope of a fully-supervised formulation for training a denoising model, which typically requires pairs of artificial mixtures and clean reference signals. Instead, a self-supervised learning approach may be better suited; this works by optimizing the model based on a pretext task which proxies the intended task \cite{SchmidhuberJ1990self}. This paradigm has seen extensive usage in computer vision research \cite{DosovitskiyA2015ExemplarCNN, DoerschC2015patch}, with even recent studies applying the concept towards speaker-agnostic speech enhancement \cite{WangYC2020selfsupervised}; our paper investigates self-supervised learning uniquely with regards to speaker-specific, thus personalized speech enhancement. 

To this end, we improve the quality of the test user's noisy data by incorporating a data purification step, as conceivably, some audio frames of the noisy speaker-specific dataset may contain more clean speech than others. Rather than treating every frame equivalently, the self-supervised formulation may benefit from additional prior knowledge which emphasizes certain frames based on the presence of clean speech. Our proposed method introduces a weighting scheme derived from a frame-by-frame estimate of the noisy speech's signal-to-noise ratio (SNR). An explicit SNR prediction step has been used before to boost the performance of DNNs for speech enhancement in a fully-supervised setup \cite{GaoT2015lvaica, FuSW2016interspeech}. However, our work is the first to apply this step in the context of personalized speech enhancement. By weighting the frames based on their SNR, we inexpensively label the unlabeled noisy data. 
This data purification can guide the speaker-specific self-supervised learning objective towards better approximating a hypothetical speaker-specific fully-supervised learning objective.

Our paper's contributions can be summarized as follows: (1) We formulate the personalized speech enhancement context, whose training is done using noisy data of the intended test-time speaker rather than the clean voice. (2) We introduce one method of self-supervised learning for personalized speech enhancement, which treats the noisy speaker data as pseudo-sources. (3) We propose a data purification step. It modifies the self-supervised learning loss function to weight the contributions of the noisy pseudo-source training data based on the frame-by-frame ``cleanliness score", or SNR.

By avoiding explicit calculation of any speaker-identifying embedding vectors, and without using any clean speech data, we assert that the proposed methods are first steps forward towards privacy-preserving personalized speech enhancement.

\section{Methods}

\subsection{Fully-Supervised Speech Enhancement}

Speech enhancement (SE) is commonly setup as a fully-supervised learning problem, in which a model learns to map noisy mixture signals to clean speech signals by processing pairs of inputs and targets. The input mixtures $\bm{x}$ are made by artificially mixing clean speech utterances $\bm{s}$ with training noise signals $\bm{n}$; the amplitude of $\bm{n}$ may be scaled to simulate various SNRs. The utterances are sampled from a large training dataset containing many speakers, $\bm{s} \in \setspkagnostic$, and the noises from a similarly large dataset of diverse noises, $\bm{n} \in \mathbb{N}$.
The denoising model $g$ updates its parameters $\mathcal{W}_g$ with each iteration such that the distance $\mathcal{E}$ between the denoised estimate signal $\bm{y}$ and the target clean speech signal $\bm{s}$ is minimized. The learning procedure for the generalist model may be summarized as follows:
\begin{align}
    \label{eq:se_mixture}\text{Mixture: }& \bm{x} = \bm{s} + \bm{n};
    \quad \bm{s} \in \setspkagnostic, \hspace{2mm} \bm{n} \in \mathbb{N} \\
    \label{eq:se_objective}\text{SE Objective: }& \operatorname*{argmin}_{\mathcal{W}_g} \hspace{2mm}
    \dist{\bm{y} = g(\bm{x}; \mathcal{W}_g)}{\bm{s}}
\end{align}

 There are many potential choices for the loss function $\mathcal{E}$---in this study, we utilize time-domain mean square error ($\mathcal{L}_{\text{MSE}}$), which is the per-sample squared distance between the estimate ($\bm{y}$) and target ($\bm{s}$) waveform pairs of length $L$,

\begin{equation}
\mathcal{L}_{\text{MSE}}(\bm{s}, \bm{y}) = \frac{1}{L} \sum_{i=0}^L \left(\bm{s}_i - \bm{y}_i \right)^{2} .
\label{eq:mse}
\end{equation}
MSE has been shown to correlate well with improving the objective signal quality \cite{KolbaekM2020ieeeacmaslp}, but denoising performance is commonly reported in scale invariant signal-to-distortion ratio (SI-SDR) \cite{LeRouxJL2018sisdr}.

A na\"ive approach to personalized speech enhancement would be to replicate this procedure using only speaker-specific data. But because we consider the personalized speech enhancement to be a no-shot learning problem, our study assumes that we do not have access to ground-truth clean speech utterances.
Therefore, the conventional fully-supervised learning objective cannot be used directly in training a personalized speech enhancement model.

\subsection{Self-Supervised (Pseudo) Speech Enhancement}

We assume that the easily collected noisy speech data from that  is an additive mixture of their clean utterances corrupted by a set of unknown additive noises, $\setspkspecificpseudo = \setspkspecific + \mathbb{M}$. We can simulate this ``premixture'' process in our experiments by sampling utterances from the test-time speaker, $\bm{s}^{(k)}  \in \setspkspecific$, and mixing them with a corpus of designated premixture noises, $\bm{m} \in \mathbb{M}$; the model is only allowed to access the premixed signals $\tilde{\bm{s}}$ and not its components $\bm{s}^{(k)}$ and $\bm{m}$. From here, we omit the speaker index, superscript $k$, for brevity.

Our proposed self-supervised learning strategy treats the premixtures as the new set of training targets. Therefore, the premixtures $\tilde{\bm{s}}$ are injected with further training noises, $\bm{n} \in \mathbb{N}$, to create a new set of input mixtures $\tilde{\bm{x}}$. The self-supervised model $f_\text{PSE}$ updates its parameters $\mathcal{W}_{f_\text{PSE}}$ by mapping the doubly-corrupted input mixtures $\tilde{\bm{x}}$ to a pseudo-denoised estimate signal $\tilde{\bm{y}}$ and minimizing its distance to the originating premixture source $\tilde{\bm{s}}$---in other words, the self-supervised model learns to undo only the second noise injection.

The two discussed mapping functions are non-equivalent, i.e. $f_\text{PSE} \neq g$, not only because $g$ is trained on data from many speakers while $f_\text{PSE}$ is trained on data from a single speaker, but also because $f_\text{PSE}$ is trained using non-clean source signals which makes it a pseudo speech enhancement (PSE) model. Suppose there exists a hypothetical optimal speaker-specific denoising function ($f^{\ast}$); our hypothesis is that $f_\text{PSE}$ better approximates $f^{\ast}$ as opposed to the fully-supervised speaker-agnostic generalist function $g$.

Because it does not directly solve the speech enhancement problem, while it still mimics the source-separating nature via data augmentation on unlabeled signals, we consider our proposed learning objective analogous to being a pretext task. The self-supervised training procedure summarized as follows:
\begin{align}
    \label{eq:pse_premixture}\text{Premixture: }& \tilde{\bm{s}} = \bm{s} + \bm{m};
    \hspace{2mm} \bm{s} \in \setspkspecific, \hspace{2mm} \bm{m} \in \mathbb{M} \\
    \label{eq:pse_mixture}\text{Mixture: }& \tilde{\bm{x}} = \tilde{\bm{s}} + \bm{n};
    \hspace{2mm} \bm{n} \in \mathbb{N} \\
    \label{eq:pse_objective}\text{PSE Objective: }& \operatorname*{argmin}_{\mathcal{W}_{f_\text{PSE}}} \hspace{2mm}
    \dist{\tilde{\bm{y}} = f_\text{PSE}(\tilde{\bm{x}}; \mathcal{W}_{f_\text{PSE}}}{\tilde{\bm{s}}}
\end{align}

\begin{figure}[t]
    \centering
    \includegraphics[width=0.9\linewidth]{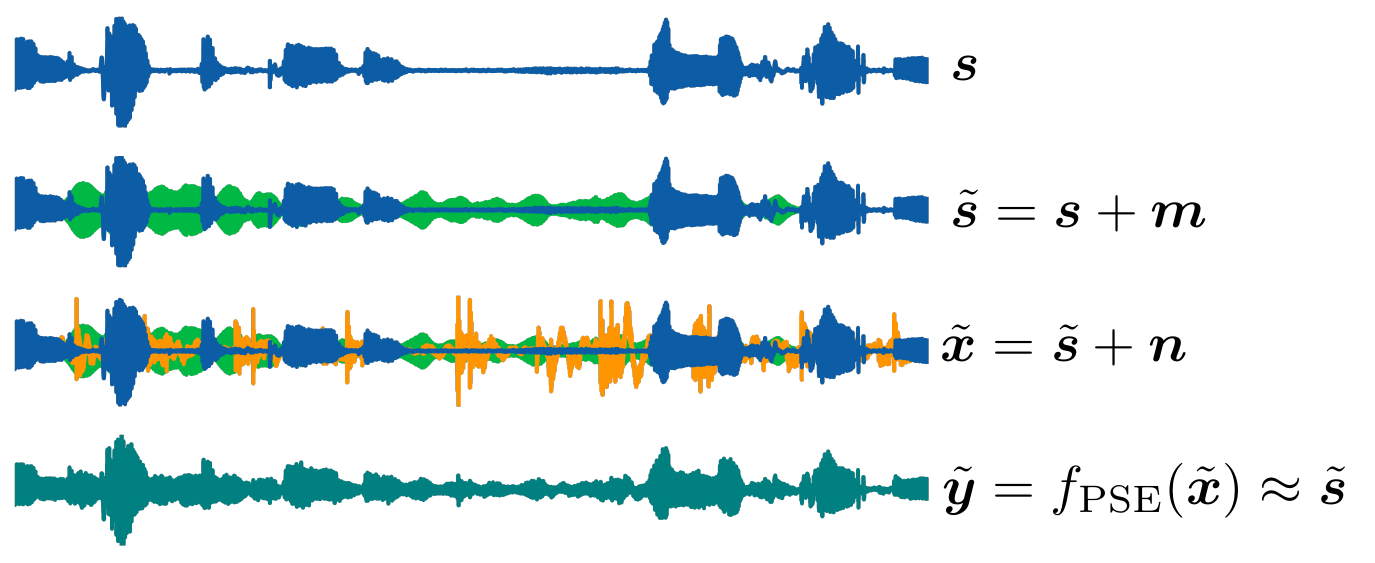}
    \caption{Illustration of the audio data transformations through pseudo speech enhancement (PSE). Premixtures $\tilde{\bm{s}}$ represent real-world noisy audio recordings from the test-time speaker, and are the training target of pseudo-denoising model $f_\text{PSE}$.}
    \label{fig:pse_waveforms}
\end{figure}

Figure \ref{fig:pse_waveforms} illustrates the impact of the two stages of noise which are applied to the clean speech waveform. In short, our specialist models trained using pseudo speech enhancement are optimized in the same manner as the generalist models, by minimizing the per-sample distance between pairs of estimates and targets, however the target in this case is a pseudo-source, i.e. $\mathcal{E}(\tilde{\bm{y}} \hspace{0.5mm} || \hspace{0.5mm} \tilde{\bm{s}}) = \mathcal{L}_{\text{MSE}}(\tilde{\bm{y}}, \tilde{\bm{s}})$.


\subsection{Data-Purified Pseudo Speech Enhancement}

The success of deriving meaningful speaker-specific features from pseudo speech enhancement depends on the quality of the premixture---more specifically, the sparsity of $\bm{m}$ in time, as well as the instantaneous SNR between $\bm{s}$ and $\bm{m}$, are both factors as to whether $\tilde{\bm{s}}$ is too degraded to be usable. If $\bm{m}$ is sufficiently sparse, some portions of the premixture may contain near-clean speech. Our goal is to inform the enhancement model of where these near-clean frames may be. We propose introducing ``data purification'' (DP) to pseudo speech enhancement training; the data-purified model $f_\text{DP}$ estimates a weighting vector $\bm{p}$ which diminishes the contribution of contaminated frames towards the loss function $\mathcal{E}$. By masking out the noisy premixture frames, the personalized model will hypothetically learn only using snippets of clean speech from the test-time user. This will differ from the original self-supervised model, i.e. $f_\text{DP} \neq f_\text{PSE}$, but our hypothesis is that it may better approximate the ideal denoising function, i.e. $f_\text{DP} \approx f^{\ast}$.

Our method for generating $\bm{p}$ is to train a separate model $h$ which can estimate the segmental, or frame-by-frame SNR of the premixtures, computed over a set of windowed overlapping frames. The SNR predictor is a regressive model trained on a diverse set of training speakers and training noises which outputs a vector of instantaneous SNRs, $\bm{\alpha}$; it has no knowledge of the test-time speaker or the test-time noise environment. Given an estimate signal $\hat{\bm{v}}$ and a target signal ${\bm{v}}$, both of length $L$, their residual is $\bm{r} = {\bm{v}} - \hat{\bm{v}}$, and the frame-by-frame/segmental SNR (SegSNR) can be defined as:
\begin{equation}
\operatorname{SegSNR}_j(\bm{v}, \hat{\bm{v}}) =
10 \log_{10}
\left[ \frac{
    \sum_{i=Hj}^{Hj+N-1} \left({w}_{i}{v}_{i}\right)^{2}
}{
    \sum_{i=Hj}^{Hj+N-1} \left({w}_{i}{r}_{i}\right)^{2}
} \right] ,
\label{eq:segmental_snr}
\end{equation}
where $N$ is the frame size, $H$ is the hop size, $j$ is a zero-based frame index (i.e. $0 \leq j \leq \ceil{\frac{L}{H}}-1$), and vector $\bm{w}$ comes from the Hann window function of length $N$. Note that the SNR predictor inputs are of length $L$ and outputs are of length $\ceil{\frac{L}{H}}$. Its training objective may then be summarized as:
\begin{align}
    \label{eq:snrp_mixture}\text{Mixture: }& \bm{x} = \bm{s} + \bm{n};
    \quad \bm{s} \in \setspkagnostic, \hspace{2mm} \bm{n} \in \mathbb{N} \\
    \label{eq:snrp_target}\text{Target: }& \bm{\alpha} = \operatorname{SegSNR}(\bm{s},\bm{x}) \\
    \label{eq:snrp_objective}\text{\shortstack[r]{SNR-Predictor\\Objective:} }& \operatorname*{argmin}_{\mathcal{W}_h} \hspace{2mm}
    \dist{\hat{\bm{\alpha}} = h(\bm{x}; \mathcal{W}_h)}{\bm{\alpha}} , 
\end{align}

When training the pseudo-denoising model, the fully-trained SNR predictor first analyses the input premixtures, $\tilde{\bm{s}} \in \setspkspecificpseudo$, to estimate the instantaneous SNRs, $\hat{\bm{\alpha}} = h(\tilde{\bm{s}})$; we apply the logistic function $\sigma$ to the $\hat{\bm{\alpha}}$ logits to obtain a set of frame-by-frame weights:
\begin{equation}
\bm{p}= \sigma(\hat{\bm{\alpha}}) = \frac{1}{1 + e^{-\hat{\bm{\alpha}}}}
\end{equation}

The training procedure for the data purified pseudo speech enhancement (PSE + DP) model mirrors Eq. \eqref{eq:pse_premixture}--\eqref{eq:pse_objective} except that we modify the loss function to now incorporate the frame-by-frame weighting vector through a custom segmental MSE function, i.e. $\dist{\tilde{\bm{y}}}{\tilde{\bm{s}}} = \mathcal{L}_{\text{SegMSE}}(\tilde{\bm{y}},\tilde{\bm{s}};\bm{p})$, where
\begin{equation}
\mathcal{L}_{\text{SegMSE}} = \frac{1}{J} \sum_{j=0}^{J\!-\!1} p_{j} \left[ \frac{1}{N}  \sum_{i=Hj}^{Hj+N-1}\!\!\left(w_i\tilde{s}_i - w_i\tilde{y}_i\right)^{2} \right].
\label{eq:segmental_mse}
\end{equation}
Here $J$ is the number of frames $\ceil{\frac{L}{H}}$. 
The mean-squared difference is taken between the windowed segments, which are then weighted by $\bm{p}$ then averaged across all frames.



\begin{figure}[t!]
    \centering
    \includegraphics[width=\linewidth]{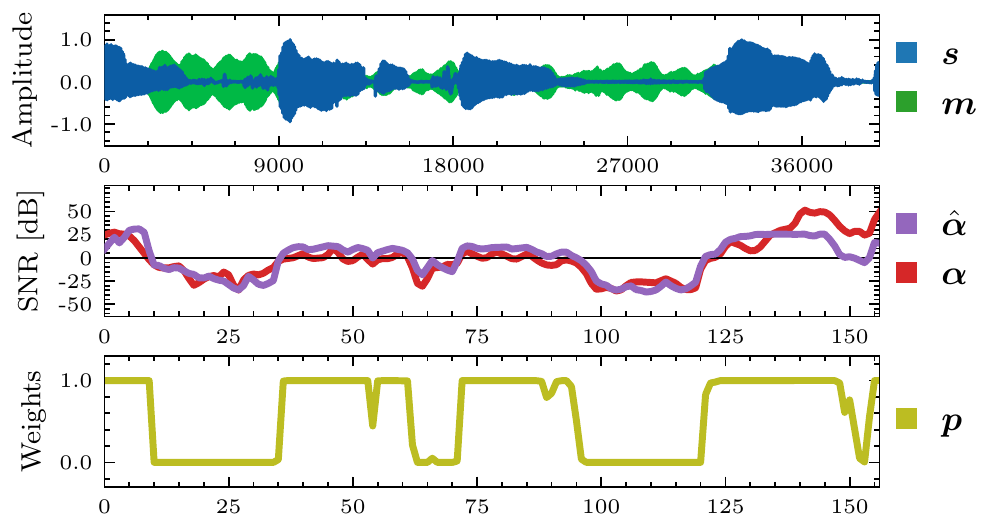}
    \caption{Illustration of the fully-trained SNR predictor inputs and outputs. The first subplot features an example premixture / pseudo-source $\tilde{\bm{s}}$. In the second subplot, the SNR predictor network $h$ estimates the instantaneous SNR of the premixture. The third subplot shows $\hat{\bm{\alpha}}$ converted to weights using the logistic function, i.e. $\bm{p} = \sigma(h(\tilde{\bm{s}}))$.}
    \label{fig:snr_prediction}
\end{figure}



\section{Experiment Setup}

\subsection{Configurations}
\label{sec:configurations}

Our experiment considers a baseline and four proposed training procedures in potentially developing personalized speech enhancement models. We group the proposed methods based on whether we pretrain the models using random initialization or speaker-agnostic fully-supervised pretraining. Regardless of initialization, all the proposed configurations use pseudo speech enhancement as the self-supervised learning approach to personalization. We additionally examine the impact of the proposed data purification scheme. 

\begin{itemize}[leftmargin=*]
    \item \textbf{SE}: Trained to minimize Eq. \eqref{eq:se_objective}. This is our generalist baseline, the speaker-agnostic speech enhancement system.
    \item \textbf{PSE}: The proposed plain pseudo speech enhancement method. This self-supervised learning method relies solely on noisy speaker-specific data to minimize Eq. \eqref{eq:pse_objective}. 
    \item \textbf{PSE+DP}: A self-supervised setup using Eq. \eqref{eq:pse_objective}. However, the model uses the weighted segmental MSE $\mathcal{L}_\text{SegMSE}$ instead Eq. \eqref{eq:segmental_mse} to purify the noisy speaker-specific dataset.  
    \item \textbf{SE$\rightarrow$PSE}: Instead of a random initialization, a model is first trained to minimize Eq. \eqref{eq:se_objective}, then fine-tuned to minimize Eq. \eqref{eq:pse_objective}.
    \item \textbf{SE$\rightarrow$PSE+DP}: Same as above, but with data purification. 
\end{itemize}

\subsection{Data Preparation} 

We opt for an online data augmentation procedure which combines three different public audio datasets (Librispeech \cite{PanayotovV2015Librispeech}, MUSAN \cite{SnyderD2015MUSAN}, and FSD50K \cite{FonsecaE2020fsd50k}) to test our methods' robustness cross-dataset. Our speaker-specific datasets $\setspkspecific$ are stochastically sampled out of all utterances from one test speaker. This done for twenty folded out test speakers. 
The speaker-agnostic dataset $\setspkagnostic$ is stochastically sampled out of all utterances across the remaining 211 speakers within the Librispeech \texttt{train-clean-100} split. The audio clips from the FSD50K \texttt{dev} split serve as the premixture noises $\mathbb{M}$. The training noises $\mathbb{N}$ come from the MUSAN \texttt{free-sound} split. Model performance is evaluated on a set of random mixtures between unseen utterances within the test-time speaker dataset combined with unseen noises from the MUSAN \texttt{sound-bible} split.

During training, for any given mixture, the sampled signals ($\bm{s}$, $\bm{m}$, $\bm{n}$) are $\SI{1}{\sec}$ in length with a sample rate of $\SI{16}{\kilo\hertz}$. Each premixture noise $\bm{m}$ is scaled uniformly at random such that the premixture SNR falls between $\si{0}$ to $\SI{15}{\decibel}$, whereas the training noises $\bm{n}$ are scaled uniformly at random such that the mixture SNR falls between $\si{-5}$ to $\SI{5}{\decibel}$. Our choice of premixture SNR range is based off of real-world scenarios, e.g. a smart speaker collecting noisy speech data in the living room.

\subsection{Models}

Our experiment focuses on the correlation between the number of effective model parameters and the improvement in speech enhancement quality (SI-SDR) obtained through the SNR-informed data purification method. Because our real-world use case relates to compute-constrained smart devices likely to perform low-latency speech enhancement on-device, we evaluate small neural network architectures which do not compete with popular speech separation models, e.g. Conv-TasNet \cite{LuoY2019conv-tasnet} or Dual-Path RNN \cite{YiL2020dualpathRNN}. Our hypothesis is that low-capacity models will benefit the most from personalized speech enhancement, as the reduced problem space might allow for a less expressive model.

We compare three two-layer GRU-based models (varying hidden units: $64$, $128$, and $256$) so as to compare the relationship between the five training configurations and increased model complexity. These models perform denoising via time-frequency masking \cite{NarayananA2013icassp}---audio waveforms are converted to the time-frequency domain using the short-time Fourier Transform (STFT) with a frame size $N=1024$ and a hop size of $H=256$; as the waveforms are $16000$ samples in length ($\SI{1}{\sec}$), this results in $J=63$ STFT frames. The denoised signal is obtained by applying the estimated time-frequency mask onto the input STFT and performing the inverse transform. The loss is computed in the time domain. 

The SNR predictor model uses a GRU with $1024$ hidden units and $3$ hidden layers; it processes STFTs with the same $N$ and $H$, performing a frame-by-frame regression. Since the predictor is only required during training, there is no incentive to minimize the number of parameters in the predictor. These choices of $N$, $H$, and $J$ apply towards Eq. \eqref{eq:segmental_snr} and \eqref{eq:segmental_mse}.

\section{Results}

\begin{figure}[t]
    \centering
    \includegraphics[width=\linewidth]{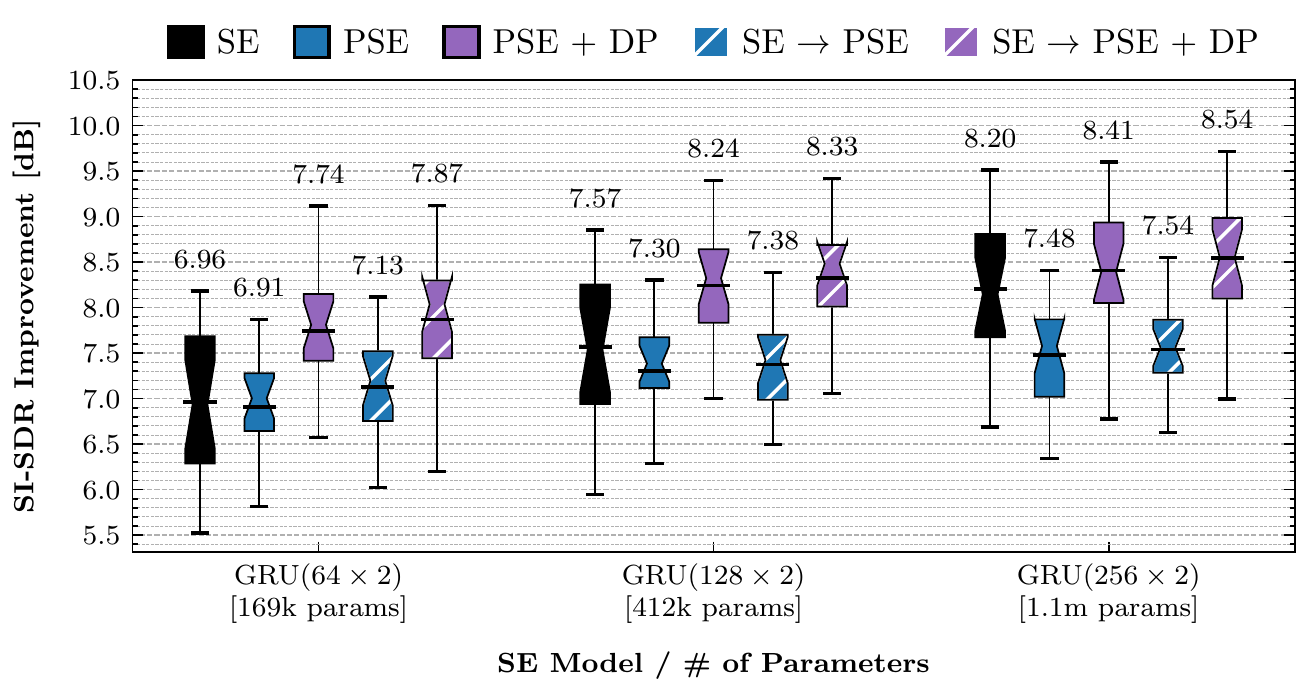}
    \caption{Box plot of experiment results. Numbers reported are the mean SI-SDR improvement on unseen mixtures across each test-time user. Model configurations are detailed in Sec. \ref{sec:configurations}.}
    \label{fig:results}
\end{figure}

Fig. \ref{fig:results} summarizes the results of our experiment, listing the mean SI-SDR improvement on unseen mixtures calculated across twenty different test-time speakers. 
First, we observe the model compression benefits of personalized speech enhancement by comparing the specialist models with fewer parameters with the generalist models with more parameters. For example, a $64$ hidden unit network, trained using \textbf{SE$\rightarrow$PSE+DP}, slightly exceeds the average denoising performance of a generalist model \textbf{SE} with 128 hidden units---this is an effective $\SI{59}{\%}$ reduction in model size (from $\SI{412}{k}$ to $\SI{169}{k}$ parameters).

The personalized models trained with the na\"ive pretext task, \textbf{PSE}, underperform compared to the baseline non-personalized \textbf{SE} models with an equivalent number of parameters. This indicates that the configuration of our experiment---a realistic premixture SNR range of $\si{0}$ to $\SI{15}{\decibel}$---challenges the self-supervised model with obtaining speech denoising features out of noisy speech. As hypothesized, data purification overcomes this difficulty by ignoring the too-noisy frames. We see across all model sizes that the data purified self-supervised models \textbf{PSE+DP} consistently outperform equivalently sized baselines.

Our experiments show that the generalist-initialized specialists outperform the randomly-initialized specialists only marginally, at most by $\SI{0.22}{\decibel}$. This suggests that the large multi-speaker corpus $\mathbb{G}$ and the model trained from it $g$ are limited in their ability to address all the peculiarities of our 20 test-time speakers. In other words, the denoising models $g$, $f_\text{PSE}$, and $f_\text{DP}$ do not approximate one another.

Intuitively, the least complex model stands to gain the most from the data purification-aided self-supervised learning. This is confirmed by our findings that the smallest personalized GRU model benefits the most, e.g., \textbf{SE$\rightarrow$PSE+DP} outperforms \textbf{SE} by $0.91$ dB, 
while the largest GRU model gains the least, e.g., \textbf{SE$\rightarrow$PSE+DP} outperforms \textbf{SE} by $0.34$ dB.

\section{Conclusion}

This work introduced personalized speech enhancement as a no-shot learning problem which motivated a self-supervised learning solution. Our method treated noisy data as pseudo-sources. We personalized speech denoising models for twenty different speakers using three neural network architectures with varied model complexity. We compared a speaker-agnostic fully-supervised model against two proposed self-supervised models: one without and another with the proposed data purification that suppresses the contribution of low-SNR frames to the learning objective. Our study showed that the smallest models benefit the most from personalization, for which the data purified self-supervised learning scheme yields the best denoising performance.

\section{Acknowledgement}

This material is based upon work supported by the National Science Foundation under Grant No. 2046963.

\newpage

\bibliographystyle{IEEEtran}

\bibliography{mjkim, local}

\begin{thebibliography}{10}
\providecommand{\url}[1]{#1}
\csname url@samestyle\endcsname
\providecommand{\newblock}{\relax}
\providecommand{\bibinfo}[2]{#2}
\providecommand{\BIBentrySTDinterwordspacing}{\spaceskip=0pt\relax}
\providecommand{\BIBentryALTinterwordstretchfactor}{4}
\providecommand{\BIBentryALTinterwordspacing}{\spaceskip=\fontdimen2\font plus
\BIBentryALTinterwordstretchfactor\fontdimen3\font minus
  \fontdimen4\font\relax}
\providecommand{\BIBforeignlanguage}[2]{{%
\expandafter\ifx\csname l@#1\endcsname\relax
\typeout{** WARNING: IEEEtran.bst: No hyphenation pattern has been}%
\typeout{** loaded for the language `#1'. Using the pattern for}%
\typeout{** the default language instead.}%
\else
\language=\csname l@#1\endcsname
\fi
#2}}
\providecommand{\BIBdecl}{\relax}
\BIBdecl

\bibitem{BollSF79ieeeassp}
S.~F. Boll, ``Suppression of acoustic noise in speech using spectral
  subtraction,'' \emph{IEEE Transactions on Acoustics, Speech, and Signal
  Processing}, vol.~27, pp. 113--120, 1979.

\bibitem{EphraimY1984spectralamplitude}
Y.~Ephraim and D.~Malah, ``Speech enhancement using a minimum-mean square error
  short-time spectral amplitude estimator,'' \emph{IEEE Transactions on
  Acoustics, Speech, and Signal Processing}, vol.~32, no.~6, pp. 1109--1121,
  1984.

\bibitem{Gannot98ieeesap}
S.~Gannot, D.~Burshtein, and E.~Weinstein, ``Iterative and sequential {Kalman}
  filter-based speech enhancement algorithms,'' \emph{IEEE Transactions on
  Speech and Audio Processing}, vol.~6, pp. 373--385, 1998.

\bibitem{XuY2014ieeespl}
Y.~Xu, J.~Du, L.-R. Dai, and C.-H. Lee, ``An experimental study on speech
  enhancement based on deep neural networks,'' \emph{IEEE Signal Processing
  Letters}, vol.~21, no.~1, pp. 65--68, 2014.

\bibitem{HuangP2015ieeeacmaslp}
P.~Huang, M.~Kim, M.~Hasegawa-Johnson, and P.~Smaragdis, ``Joint optimization
  of masks and deep recurrent neural networks for monaural source separation,''
  \emph{IEEE/ACM Transactions on Audio, Speech, and Language Processing},
  vol.~23, no.~12, pp. 2136--2147, Dec 2015.

\bibitem{WeningerF2015lvaica}
F.~Weninger, H.~Erdogan, S.~Watanabe, E.~Vincent, J.~{Le Roux}, J.~R. Hershey,
  and B.~Schuller, ``{Speech Enhancement with LSTM Recurrent Neural Networks
  and its Application to Noise-Robust ASR},'' in \emph{Proceedings of the
  International Conference on Latent Variable Analysis and Signal Separation
  (LVA/ICA)}, Aug. 2015.

\bibitem{PascualS2017segan}
S.~Pascual, A.~Bonafonte, and J.~Serra, ``Segan: Speech enhancement generative
  adversarial network,'' \emph{arXiv preprint arXiv:1703.09452}, 2017.

\bibitem{WangDL2018ieeeacmaslp}
D.~L. {Wang} and J.~{Chen}, ``Supervised speech separation based on deep
  learning: An overview,'' \emph{IEEE/ACM Transactions on Audio, Speech, and
  Language Processing}, vol.~26, no.~10, pp. 1702--1726, 2018.

\bibitem{ChenJ2016largescale}
{J. Chen and Y. Wang and S. E. Yoho and D. Wang and E. W. Healy}, ``Large-scale
  training to increase speech intelligibility for hearing-impaired listeners in
  novel noises,'' \emph{Journal of the Acoustical Society of America}, vol.
  139, no.~5, pp. 2604--2612, 2016.

\bibitem{ChenJ2017generalization}
{J. Chen and D. Wang}, ``Long short-term memory for speaker generalization in
  supervised speech separation,'' \emph{Journal of the Acoustical Society of
  America}, vol. 141, no.~6, pp. 4705--4714, 2017.

\bibitem{LiuD2014interspeech}
D.~Liu, P.~Smaragdis, and M.~Kim, ``Experiments on deep learning for speech
  denoising,'' in \emph{Proceedings of the Annual Conference of the
  International Speech Communication Association (Interspeech)}, Sep 2014.

\bibitem{WangQ2018voicefilter}
Q.~Wang, H.~Muckenhirn, K.~Wilson, P.~Sridhar, Z.~Wu, J.~Hershey, R.~A.
  Saurous, R.~J. Weiss, Y.~Jia, and I.~L. Moreno, ``Voicefilter: Targeted voice
  separation by speaker-conditioned spectrogram masking,'' \emph{arXiv preprint
  arXiv:1810.04826}, 2018.

\bibitem{KolbakM2017ieeeacmaslp}
M.~Kolb{\ae}k, Z.~H. Tan, and J.~Jensen, ``Speech intelligibility potential of
  general and specialized deep neural network based speech enhancement
  systems,'' \emph{IEEE/ACM Transactions on Audio, Speech, and Language
  Processing}, vol.~25, no.~1, pp. 153--167, Jan 2017.

\bibitem{SivaramanA2020interspeech}
A.~Sivaraman and M.~Kim, ``{Sparse Mixture of Local Experts for Efficient
  Speech Enhancement},'' in \emph{Proceedings of the Annual Conference of the
  International Speech Communication Association (Interspeech)}, 2020, pp.
  4526--4530.

\bibitem{rochford2019accessibility}
J.~Rochford, ``{Accessibility and IoT / Smart and Connected Communities},''
  \emph{AIS Transactions on Human-Computer Interaction}, vol.~11, no.~4, pp.
  253--263, 2019.

\bibitem{JiaY2018neurips}
Y.~Jia, Y.~Zhang, R.~Weiss, Q.~Wang, J.~Shen, F.~Ren, Z.~Chen, P.~Nguyen,
  R.~Pang, I.~L. Moreno, and Y.~Wu, ``{Transfer Learning from Speaker
  Verification to Multispeaker Text-To-Speech Synthesis},'' in \emph{Advances
  in Neural Information Processing Systems (NeurIPS)}, vol.~31, 2018.

\bibitem{ChaoWL2016zeroshotlearning}
{W.-L. Chao, and S. Changpinyo, and B. Gong, and F. Sha}, ``{An Empirical Study
  and Analysis of Generalized Zero-Shot Learning for Object Recognition in the
  Wild},'' in \emph{Proceedings of the European Conference on Computer Vision
  (ECCV)}, 2016, pp. 52--68.

\bibitem{XianY2018zeroshotlearning}
{Y. Xian and C. H. Lampert and B. Schiele and Z. Akata}, ``{Zero-Shot Learning
  -- A Comprehensive Evaluation of the Good, the Bad and the Ugly},''
  \emph{IEEE Transactions on Pattern Analysis and Machine Intelligence},
  vol.~41, no.~9, pp. 2251--2265, 2018.

\bibitem{SchmidhuberJ1990self}
J.~Schmidhuber, ``Making the world differentiable: On using self-supervised
  fully recurrent neural networks for dynamic reinforcement learning and
  planning in non-stationary environments,'' 1990.

\bibitem{DosovitskiyA2015ExemplarCNN}
A.~Dosovitskiy, P.~Fischer, J.~T. Springenberg, M.~Riedmiller, and T.~Brox,
  ``{Discriminative Unsupervised Feature Learning with Exemplar Convolutional
  Neural Networks},'' \emph{arXiv preprint arXiv:1406.6909}, 2015.

\bibitem{DoerschC2015patch}
C.~Doersch, A.~Gupta, and A.~A. Efros, ``{Unsupervised Visual Representation
  Learning by Context Prediction},'' in \emph{Proceedings of the International
  Conference on Computer Vision (ICCV)}, 2015.

\bibitem{WangYC2020selfsupervised}
Y.-C. Wang, S.~Venkataramani, and P.~Smaragdis, ``{Self-supervised Learning for
  Speech Enhancement},'' \emph{arXiv preprint arXiv:2006.10388}, 2020.

\bibitem{GaoT2015lvaica}
T.~Gao, J.~Du, Y.~Xu, C.~Liu, L.-R. Dai, and C.-H. Lee, ``{Improving Deep
  Neural Network Based Speech Enhancement in Low SNR Environments},'' in
  \emph{Proceedings of the International Conference on Latent Variable Analysis
  and Signal Separation (LVA/ICA)}, 2015, pp. 75--82.

\bibitem{FuSW2016interspeech}
S.~W. Fu, Y.~Tsao, and X.~Lu, ``{SNR-Aware Convolutional Neural Network
  Modeling for Speech Enhancement},'' in \emph{Proceedings of the Annual
  Conference of the International Speech Communication Association
  (Interspeech)}, 2016, pp. 3768--3772.

\bibitem{KolbaekM2020ieeeacmaslp}
M.~Kolb{\ae}k, Z.-H. Tan, S.~H. Jensen, and J.~Jensen, ``{On Loss Functions for
  Supervised Monaural Time-Domain Speech Enhancement},'' \emph{IEEE/ACM
  Transactions on Audio, Speech, and Language Processing}, vol.~28, pp.
  825--838, 2020.

\bibitem{LeRouxJL2018sisdr}
J.~L. Roux, S.~Wisdom, H.~Erdogan, and J.~R. Hershey, ``{SDR} - half-baked or
  well done?'' \emph{arXiv preprint arXiv:1811.02508}, 2018.

\bibitem{PanayotovV2015Librispeech}
V.~Panayotov, G.~Chen, D.~Povey, and S.~Khudanpur, ``{Librispeech}: {An} {ASR}
  corpus based on public domain audio books,'' in \emph{Proceedings of the IEEE
  International Conference on Acoustics, Speech, and Signal Processing
  (ICASSP)}.\hskip 1em plus 0.5em minus 0.4em\relax IEEE, 2015, pp. 5206--5210.

\bibitem{SnyderD2015MUSAN}
D.~Snyder, G.~Chen, and D.~Povey, ``{MUSAN}: {A Music, Speech, and Noise
  Corpus},'' \emph{arXiv preprint arXiv:1510.08484}, 2015.

\bibitem{FonsecaE2020fsd50k}
E.~Fonseca, X.~Favory, J.~Pons, F.~Font, and X.~Serra, ``{FSD50K: an Open
  Dataset of Human-Labeled Sound Events},'' \emph{arXiv preprint
  arXiv:2010.00475}, 2020.

\bibitem{LuoY2019conv-tasnet}
Y.~Luo and N.~Mesgarani, ``{Conv-TasNet}: Surpassing ideal time--frequency
  magnitude masking for speech separation,'' \emph{IEEE/ACM Transactions on
  Audio, Speech, and Language Processing}, vol.~27, no.~8, pp. 1256--1266,
  2019.

\bibitem{YiL2020dualpathRNN}
Y.~Luo, Z.~Chen, and T.~Yoshioka, ``{Dual-path RNN}: efficient long sequence
  modeling for time-domain single-channel speech separation,'' in
  \emph{Proceedings of the IEEE International Conference on Acoustics, Speech,
  and Signal Processing (ICASSP)}, 2020.

\bibitem{NarayananA2013icassp}
A.~Narayanan and D.~L. Wang, ``Ideal ratio mask estimation using deep neural
  networks for robust speech recognition,'' in \emph{Proceedings of the IEEE
  International Conference on Acoustics, Speech, and Signal Processing
  (ICASSP)}, May 2013, pp. 7092--7096.

\end{thebibliography}

\end{document}